\documentclass[letterpaper]{jpconf}

\usepackage{amssymb}
\usepackage{amsthm}
\usepackage{graphicx}
\usepackage{bm}
\newcommand{\Rea}{\mbox{Re}}

\begin{document}

%LA-UR-06-8374

\title{Decohering histories and open quantum systems}

\author{Eric D. Chisolm}

\address{Theoretical Division, Los Alamos National Laboratory, Los Alamos, 
         NM~~87545}

\ead{echisolm@lanl.gov}

\begin{abstract}
I briefly review the ``decohering histories'' or ``consistent histories''
formulation of quantum theory, due to Griffiths, Omn\`{e}s, and
Gell-Mann and Hartle (and the subject of my graduate work with George
Sudarshan).  I also sift through the many meanings that have been
attached to decohering histories, with an emphasis on the most basic
one: Decoherence of appropriate histories is needed to establish that
quantum mechanics has the correct classical limit.  Then I will
describe efforts to find physical mechanisms that do this.  Since most
work has focused on density matrix versions of decoherence, I'll
consider the relation between the two formulations, which historically has
not been straightforward.  Finally, I'll suggest a line of research
that would use recent results by Sudarshan to illuminate this aspect
of the classical limit of quantum theory.
\end{abstract}

\section{Introduction}

Many of the most interesting developments in quantum theory since
Bell's Theorem have centered on the notion of ``decoherence,''
understood most generally as the absence of interference effects.
Decoherence is to be avoided at all costs when constructing mesoscopic
or macroscopic quantum devices, such as SQUIDs or quantum computers,
while its unavoidability in certain domains is said to be an important
part of why large parts of the world appear classical.  One area where
decoherence has played an important role is the so-called ``decohering
histories'' or ``consistent histories'' formulation of quantum theory,
due to Griffiths \cite{griffiths}, Omn\`{e}s \cite{omnes}, and
Gell-Mann and Hartle \cite{GMH}.  While it has not been adopted
universally as an interpretive framework for quantum mechanics, which
is what it was originally proposed for, it is certainly a useful tool
for clarifying certain aspects of the classical limit of quantum
theory, which remains a challenging problem despite the enormous
progress made in the last few decades.  In this article I will review
the basic notions underlying decohering histories, and I will also
spend a little time discussing various interpretations of this
formalism.  Having settled for application to the classical limit,
leaving questions of interpretation aside, I will then compare the
decohering histories approach to decoherence as seen in the time
evolution of reduced density matrices.  Since I argue that
demonstrating decoherence of appropriate histories is needed to
properly establish the classical limit of quantum theory, connecting
these two approaches is of some importance.  I will run across a
problem which Sudarshan and I discussed at some length near the end of
my time with him, without our coming to any good solution.  Finally, I
will apply recent results due to Sudarshan and coworkers to suggest a
potentially very useful resolution of the issue, one that could
illuminate not just this area but other aspects of the classical limit
of quantum mechanics.

\section{Histories and decoherence}

The formalism of quantum mechanics is related to empirical
results by the \emph{Born rule}, which states the following: If a
system is in state $|\Psi\rangle$ at time $0$, then the probability at
time $t$ that the system will be found in state $|\alpha\rangle$ is
given by
\begin{equation}
p(\alpha) = \left| \langle \alpha | U(t) | \Psi \rangle \right|^2 
\label{pureBorn}
\end{equation}
where $U(t)$ is the time development operator which evolves the state
from time $0$ to time $t$.  This expression can be rewritten 
\begin{eqnarray}
p(\alpha) & = & \Tr[P_\alpha(t) \rho] \nonumber \\
          & = & \Tr[P_\alpha(t) \rho P_\alpha(t)],
\label{genBorn}
\end{eqnarray}
where $\rho = | \Psi \rangle \langle \Psi |$ is the initial density
operator of the system, $P_\alpha$ is the projection operator onto
state $| \alpha \rangle$, and I have switched to the Heisenberg
picture, in which observables evolve in time but states do not.  The
second line follows from the first because projection operators are
idempotent and the trace is cyclic in its arguments.  This expression
is actually more general than Eq.\ (\ref{pureBorn}) because it remains
valid when $\rho$ is a general density operator, representing an
incoherent statistical mixture of pure states given by projectors.

An obvious generalization of this rule gives the probability not for a
single measurement result, but for a series of measurement results at
a succession of times:
\begin{equation}
p(\alpha_1, \alpha_2, \ldots, \alpha_n) = \Tr[P_{\alpha_n}(t_n) \cdots 
              P_{\alpha_2}(t_2) P_{\alpha_1}(t_1) \rho P_{\alpha_1}(t_1) 
              P_{\alpha_2}(t_2) \cdots P_{\alpha_n}(t_n)].
\label{manyeventBorn}
\end{equation}
This rule was given by Wigner \cite{wig}, and Aharanov, Bergmann, and
Lebowitz discussed both this rule and its time-symmetric
generalization \cite{ABL}, but it is best known now because of its use
in describing decohering histories.  Before I get to decoherence,
though, let me introduce some simplifying notation.  Let $\alpha$
denote the sequence $\alpha_1, \alpha_2, \ldots, \alpha_n$, and given
a sequence of measurements as above I define the \emph{history}
$C_\alpha$ by
\begin{equation}
C_\alpha = P_{\alpha_1}(t_1) P_{\alpha_2}(t_2) \cdots P_{\alpha_n}(t_n),
\end{equation}
in terms of which Eq.\ (\ref{manyeventBorn}) becomes the much tidier result
\begin{equation}
p(\alpha) = \Tr(C_\alpha^\dag \rho C_\alpha).
\label{histprob}
\end{equation}
This expression allows for much more conceptual contact between
quantum and classical theory, as a history is the quantum mechanical
analog of a trajectory of a classical particle.  We routinely speak of
such a particle occupying a succession of positions at a series of
times, perhaps as it orbits another body, and in standard expositions
of quantum theory it is not obvious what the corresponding quantum
object is.  This immediately suggests that histories may be of use in
elucidating the classical limit of quantum mechanics, and that hope
will be borne out below.

Since histories are products of projection operators, they inherit
some (but not all) of the properties of their factors.  Projection
operators come in sets that satisfy
\begin{eqnarray}
P_\alpha P_\beta & = & P_\alpha \delta_{\alpha \beta} \nonumber \\
\sum P_\alpha & = & \mathbb{I}
\label{projcomplete}
\end{eqnarray}
where $\mathbb{I}$ is the identity operator.  The first condition is
that the set $\{P_\alpha\}$ is \emph{exclusive}; measurement results
corresponding to two different $P_\alpha$ cannot occur simultaneously.
[This restriction is lifted when one replaces projection operators
with positive operator valued measures (POVM), but as I'll show in a
moment exclusivity is important for defining decoherence, so how one
would construct decohering histories using POVM is not clear.]  The
second condition guarantees that $\{P_\alpha\}$ is \emph{exhaustive},
so every possible measurement result is included.  Given a set of
projection operators $\{P_{\alpha_i}\}$ at each of a predetermined set
of times $t_1, t_2, \ldots, t_n$, one can construct a set of histories
$\{C_\alpha\}$ that satisfy
\begin{equation} \sum C_\alpha = \mathbb{I}, \label{histcomplete} \end{equation}
so the histories are also exhaustive.  Since a given set of histories
defines a complete statistical sample space for a system, it makes
sense to ask whether the probabilities associated with the histories
by Eq.\ (\ref{histprob}) satisfy the classical Kolmogorov axioms of
probability:
\renewcommand{\theenumi}{\arabic{enumi}}
\begin{enumerate}
\item $p(\alpha) \geq 0$ for all $\alpha$.
\item $\sum_\alpha p(\alpha) = 1$.
\item $p(\alpha \mbox{\ or } \beta) = p(\alpha) + p(\beta)$ for all $\alpha \neq \beta$.
\end{enumerate}
Axiom $1$ is clearly satisfied by Eq.\ (\ref{histprob}), and Axiom $2$
is satisfied thanks to Eqs.\ (\ref{projcomplete}) and
(\ref{histcomplete}), but we're in trouble with Axiom $3$.  This is
easy to see if we recall that composite measurements in quantum theory
are defined by taking the sum of the corresponding projection
operators; e.g.\ if $P_1$ represents the spin of a spin-$1$ system
being found up in the $z$ direction, and $P_{-1}$ represents the spin
being found down, then
\begin{equation} P = P_1 + P_{-1} \label{addproj} \end{equation}
represents the spin being found to be nonzero.  If we apply this same
logic to histories then the operator representing either history
$\alpha$ or history $\beta$ is given by
\begin{equation} C_{\alpha \mbox{\ or } \beta} = C_\alpha + C_\beta. \end{equation}
(Notice that this is obviously correct if the two histories have the
same events at all times but one, by the remarks that led to Eq.\
(\ref{addproj}).)  Therefore
\begin{eqnarray}
p(\alpha \mbox{\ or } \beta) & = & \Tr(C_{\alpha \mbox{\ or } \beta}^\dag \, \rho \, 
                                       C_{\alpha \mbox{\ or } \beta}) \nonumber \\
       & = & \Tr(C_\alpha^\dag \rho C_\alpha) + \Tr(C_\beta^\dag \rho C_\beta) + 
             \Tr(C_\alpha^\dag \rho C_\beta) + \Tr(C_\beta^\dag \rho C_\alpha) \nonumber \\
       & = & p(\alpha) + p(\beta) + 2 \Rea\Tr(C_\alpha^\dag \rho C_\beta),
\end{eqnarray}
and Axiom $3$ holds for all histories if and only if the final term
vanishes for any $\alpha \neq \beta$.  Now we recognize this term; it
is nothing more than a generalization of the \emph{interference} terms
we see ubiquitously in quantum theory.  In fact, if we were to take a
two-slit interferometer as our system and consider two-event histories
that consisted of a particle passing through one of the two slits and
then striking the screen, this term would be exactly the usual
interference term.  Thus we have generalized not only the Born rule
but the expression for interference, and we have seen that in general
it is the presence of interference that prevents quantum mechanical
probabilities from satisfying the classical axioms of probability.

All of this inspires some new definitions.  Given a complete set of
histories, I define the \emph{decoherence functional} between any two
of them to be
\begin{equation} D(\alpha, \beta) = \Tr(C_\alpha^\dag \rho C_\beta). \end{equation}
Now $D(\alpha, \alpha)$ is the probability that history $\alpha$
occurs (cf.\ Eq.\ (\ref{histprob})), and if $\alpha \neq \beta$ then
the real part of $D(\alpha, \beta)$ is the interference term between
histories $\alpha$ and $\beta$.  Thus a set of histories obeys the
Kolmogorov rules of probability if and only if
\begin{equation}
\Rea D(\alpha, \beta) = p(\alpha) \delta_{\alpha\beta}.
\label{weakdeco}
\end{equation} 
This condition is called \emph{weak decoherence}.  In many practical
calculations the stronger condition
\begin{equation}
D(\alpha, \beta) = p(\alpha) \delta_{\alpha\beta}.
\label{meddeco}
\end{equation} 
is actually achieved; this is called \emph{medium decoherence}.
(These names suggest that there ought to be some notion of strong
decoherence, and several definitions have been proposed, but none have
caught on, so I won't bother describing them.)  Exactly what
decoherence means physically is the subject of the next section.

\section{Interpretation}

When decoherence of histories was introduced by Griffiths
\cite{griffiths}, who called it \emph{consistency} of histories, his
goal was to develop a language for speaking meaningfully about the
dynamics of a \emph{closed} quantum system, for which the Copenhagen
interpretation has nothing to say because a closed system by
definition is not being measured.  His idea was to allow the
mathematical formalism of the theory tell him how to speak
meaningfully about its dynamics; he derived Eq.\ (\ref{histprob}) for
the probability of a history, and he concluded that the formalism
allows only those sets of histories which obey the Kolmogorov axioms
to be assigned probabilities in a meaningful way.  Thus he was led to
Eq.\ (\ref{weakdeco}), and he argued that if a set of histories of a
closed quantum system exhibits weak decoherence, then it is meaningful
to say that one of the histories in question actually happens in the
system, with probabilities given by Eq.\ (\ref{histprob}).  Soon
after, Omn\`{e}s \cite{omnes} arrived at similar expressions but with
more of a focus on the logic of quantum theory, and he later
integrated them into his own interpretation of quantum mechanics,
which he describes as ``a consistent and complete reformulation of the
Copenhagen interpretation'' which lacks Copenhagen's flaws
\cite{revmodphys}.  In a slightly different vein, Gell-Mann and Hartle
\cite{GMH} had their sights set on quantum cosmology, where it had
been realized that the Copenhagen interpretation was singularly
inappropriate, the universe as a whole being the very model of a
system with no outside observers.  They have developed a quantum
theory of closed systems, in which decoherence (they first applied
this term to histories) is the sieve that determines \`{a} la
Griffiths which histories are candidates for reality and which are
not.  Various objections have been raised to the use of decohering
histories in this fashion, most famously by Dowker and Kent
\cite{dowkent}, and less famously by Sudarshan, Jordan, and me at
about the same time \cite{us}, which elicited thorough responses from
Griffiths \cite{choice} that were further elaborated in his extensive
development of his program in \cite{grifbook}.

In contrast, in this paper I would like to focus less on
interpretation and more on the relevance of decohering histories to
the classical limit.  Roughly speaking, the classical limit of quantum
mechanics consists of two components: Classical statistics and
classical dynamics.  By classical statistics, I mean that the
probabilities associated with histories in classical physics obey the
Kolmogorov rules of probability given above; Axiom 3 in particular is
an essential part of our classical intuition, from coin tossing to
weather prediction.  By classical dynamics, I mean that the
probabilities assigned by theory are strongly peaked around histories
that correspond to solutions of the classical equations of motion; it
is this aspect of the classical limit that received the lion's share
of attention until about 30 years ago, when decoherence in various
forms made its appearance.  Both components, however, are equally
necessary, which means that decoherence of an appropriately chosen set
of histories is a precondition for a quantum system to have a sensible
classical limit, and it is in precisely those histories that classical
behavior will be seen.  Thus even if all we care about is establishing
classical behavior for all practical purposes (precisely John Bell's
famous FAPP \cite{bell}), decoherence of appropriate histories must be
demonstrated.  Now I'll describe how this has been attempted.

\section{Mechanisms of decoherence}

By far the most popular option in the literature for achieving decoherence 
is to separate the quantum system under consideration into
a preferred subsystem and the remainder, often called the
\emph{environment}.  Such a separation corresponds to an expression of
the system's Hilbert space $\mathcal{H}$ as
\begin{equation} \mathcal{H} = \mathcal{H^S} \otimes \mathcal{H^E}, \end{equation} 
where $\mathcal{H^S}$ is the Hilbert space of the subsystem and
$\mathcal{H^E}$ describes the environment.  This decomposition is
actually quite general, and it corresponds to the expression of the
configuration space of a classical system as the Cartesian product of
subspaces carrying different degrees of freedom; for example, a single
particle with spin moving in space can be described as the product of
two subsystems, namely the spatial and spin degrees of freedom, each
with its own Hilbert space.  Once this separation is defined,
decoherence is achieved by considering only histories composed of
projections $P^\mathcal{S}$ onto the subsystem and by allowing the
subsystem and environment to interact, which generally greatly
accelerates the process of decoherence.  This second factor leads to
the use of open systems theory for decoherence, as the time evolution
of the subsystem is normally far from Hamiltonian even if a
Hamiltonian describes the full system.

Some of the most popular model systems for illustrating decoherence,
as described by Stamp \cite{stamp}, can be classified as follows.  The
first and most famous are harmonic oscillators coupled linearly to
baths of similar oscillators, typically in a thermal state with some
simplified spectrum of frequencies; the preferred oscillator exhibits
decoherence in its position basis due to interactions with the bath.
The influence functional technique, due to Feynman and Vernon
\cite{feynvern}, was applied to this system by Caldeira and Leggett
\cite{caldleg}, and their calculation is still widely cited as an
illustration of the efficiency with which environmental interaction
produces decoherence.  Caldeira and Leggett were at some pains to
argue in their work that this model was considerably more generic than
it would seem at first sight; the need for such arguments should be
clear, given that their intent is to illustrate what should be one of
the most widespread phenomena in nature (responsible for the entire
classical world, after all).  A second set of calculations has focused
on finite-dimensional quantum systems coupled to spin baths
\cite{prosta}; the rationale was that such systems should do a good
job of representing the low-energy dynamics of condensed matter
systems, in which generally only a few low-lying energy levels should
be excited.  A third mechanism is ``third-party decoherence''
\cite{stamp}, in which correlations between subsystems which suppress
interference arise not because of any interaction between them, but
due to their mutual interactions with a third subsystem.  (The
particular example Stamp gives is of a particle passing through a
two-slit interferometer which loses spatial interference because its
spin interacts with the slits, even though its spin and spatial
degrees of freedom are uncoupled.)

These sorts of calculations have two features in common which are
worth mentioning.  First, almost all of them are performed not on
decoherence functionals, but on \emph{reduced density matrices}
describing only the preferred subsystem; for example, Caldeira and
Leggett actually showed that the reduced density matrix of the
preferred oscillator diagonalizes rapidly in the position
representation for a variety of initial states, while its diagonal
elements remain relatively unchanged.  There have been exceptions,
such as \cite{dowhal}, but the majority of calculations in the
literature have considered not histories but reduced density matrices or
related objects.  Given that I have argued for the importance of decoherence
functionals in establishing classicality, it is worth asking how much
light density matrix calculations shed on decohering histories.
Second, a large fraction of the calculations have either assumed, or made
other simplifying assumptions to guarantee, that the subsystem's evolution
is described by a master equation of some sort, often Markovian.  When 
Sudarshan and I discussed this state of affairs, he expressed his concern
that such approximations were far more restrictive than these authors perhaps
realized, and that they cast some doubt on their claims to have illustrated
generic features that give rise to the classical world.  These issues are the
subject of the next two sections.

\section{Decoherence functionals vs.\ reduced density matrices}

Particularly since the histories approach to decoherence puts so much
emphasis on diagonalizing a functional, it is easy to misunderstand
the importance of diagonalizing reduced density matrices in that
approach (at least it was easy for me to misunderstand it).  Every
density matrix, reduced or not, is of course always diagonal in some
basis simply because it is self-adjoint.  The goal of the reduced
density matrix approach is this: Given a full system with initial
state $\rho$, reduced initial state $\Tr_\mathcal{E}\rho$, and full
time evolution operator $U(t)$, the time evolution of the reduced
density matrix
\begin{equation}
\Tr_\mathcal{E}\rho \rightarrow \Tr_\mathcal{E}[U(t) \rho U^\dag(t)]
\label{redtimeevolve}
\end{equation}
sometimes has the property that the result diagonalizes rapidly in a
particular basis, far more rapidly than the diagonal elements in that
basis undergo change, and it stays diagonal over time.  Further, this
behavior is largely independent of the initial state $\rho$.  The
states making up such a basis, possibly with additional properties,
are called ``pointer states'' \cite{zurek}, and the idea behind the
reduced density matrix approach to decoherence is that such bases
serve to define a preferred class of observables in which classical
behavior emerges in appropriate systems.  This is actually a very
sensible idea, and in fact it deals directly with the fundamental
issue raised by Dowker and Kent \cite{dowkent}, which is that in
general decoherence of histories depends far too sensitively on the
exact choice of projection operators in the histories and the times at
which they are placed.  If anything is obvious about classicality, it
is that it's robust; one does not have to look at the positions of a
planet orbiting the sun at a carefully orchested sequence of times in
order to see different alternate trajectories failing to interfere.
Thus the obvious question to ask is whether histories composed of
projections onto pointer states decohere.  To see whether this is so,
let me consider a representative decoherence functional
\begin{equation}
D(\alpha,\beta) = \Tr\left[P^\mathcal{S}_{\alpha_n}(t_n) \cdots 
                  P^\mathcal{S}_{\alpha_1}(t_1) \rho P^\mathcal{S}_{\beta_1}(t_1) 
                  \cdots P^\mathcal{S}_{\beta_n}(t_n)\right].
\end{equation}
Let me also make this expression a bit more explicit by expressing the
trace as the composition of two partial traces, over the environment
and subsystem, and let me also leave the Heisenberg picture and write
the time evolutions explicitly, with the result
\begin{equation} 
D(\alpha, \beta) = \Tr_\mathcal{S}\Tr_\mathcal{E}\big[
                   P^\mathcal{S}_{\alpha_n} U(t_n,t_{n-1}) \cdots 
                   P^\mathcal{S}_{\alpha_1} U(t_1,t_0) \rho 
                   U(t_0,t_1) P^\mathcal{S}_{\beta_1} \cdots U(t_{n-1},t_n) 
                   P^\mathcal{S}_{\beta_n}\big].
\label{opendecofunc}
\end{equation}
Now let us suppose that the time evolution in Eq.\
(\ref{redtimeevolve}) does in fact diagonalize the reduced density
matrix in some basis, and further that the projections
$P^\mathcal{S}_{\alpha_i}$ project onto that very basis.  Can we
conclude that this functional (or at least its real part) vanishes if
any $\alpha_i \neq \beta_i$?  Not obviously, for the simple reason
that the time evolution is repeatedly interrupted by projections
before the partial trace over the environment is taken, and while the
partial trace commutes with those projections, it does not commute
with the time evolution.  This is all obvious, of course, and this
problem is discussed in some detail by Kiefer in Chapter 5 of
\cite{joos}.  An important reference in that book is work by Paz and
Zurek \cite{paz}, whose solution, with notation modified for
consistency with other work shown below, is as follows.  Suppose there
exist operators $M(t_2,t_1)$ such that for any operator $A$ acting on
the full system,
\begin{equation} 
\Tr_\mathcal{E}[U(t_2,t_1) A U(t_1,t_2)] = M(t_2,t_1) \{\Tr_\mathcal{E} A\}; 
\label{evolveA}
\end{equation}
in other words, suppose that the time evolution of Eq.\ (\ref{redtimeevolve})
can be expressed as an operator acting on the reduced state only.  In that
case Eq.\ (\ref{opendecofunc}) can be written
\begin{equation}
D(\alpha,\beta) = \Tr_\mathcal{S}\big[P^\mathcal{S}_{\alpha_n} 
                  M(t_n,t_{n-1})\big\{ \cdots P^\mathcal{S}_{\alpha_1} 
                  M(t_1,t_0) \{ \Tr_\mathcal{E} \rho \} P^\mathcal{S}_{\beta_1} 
                  \cdots \big\} P^\mathcal{S}_{\beta_n}\big]. 
\end{equation}
Now it's obvious that if the times $t_i$ are sufficiently well separated for
the decoherence mechanism to do its work, the result of each $M(t_i, t_{i-1})$ will
be diagonal in the pointer basis and the $\alpha_i \neq \beta_i$ terms will vanish.
But is it reasonable to hypothesize such an operator $M$?  For some time I struggled
with this issue, because the justification offered by Paz and Zurek seemed
to rely excessively on special assumptions, such as a Markovian master equation, of
the type with which Sudarshan expressed such concern.  However, recent work by 
Sudarshan and coworkers has resolved this issue rather neatly, which brings me to
my final topic.

\section{Decoherence functionals and the theory of open systems}

Despite all my concerns with master equations and such, it was shown recently by
Jordan, Shaji, and Sudarshan \cite{jorsha} that time evolution of operators in
a subsystem of a full quantum system experiencing unitary evolution can always
be expressed in a form very similiar to Eq.\ (\ref{evolveA}) with no assumptions 
whatsoever.  Suppressing the time indices for the moment for convenience, the 
expression is
\begin{equation} M(\Tr_\mathcal{E}A) = L(\Tr_\mathcal{E}A) + K \end{equation}
where
\begin{equation}
L(\Tr_\mathcal{E}A) = \Tr_\mathcal{E}\left[U \left(\Tr_\mathcal{E}A 
                      \otimes \frac{\mathbb{I}^\mathcal{E}}
                      {\dim \mathcal{H^E}} \right) U^\dag\right]
\label{linpart}
\end{equation}
and 
\begin{equation}
K = \Tr_\mathcal{E}\left[U\left(A - \Tr_\mathcal{E}A \otimes 
    \frac{\mathbb{I}^\mathcal{E}}{\dim \mathcal{H^E}}\right)U^\dag\right],
\label{affpart}
\end{equation}
where I have reexpressed their results using my own notation.  Several
comments are in order concerning this form.  First, the whole thing
may look like simple sleight of hand (after all, this is simply Eq.\
(\ref{redtimeevolve}) with Eq.\ (\ref{linpart}) added and subtracted),
and it would be if this form had no advantages; but Jordan \textit{et
al.}\ prove several significant results about this form and develop
useful expressions for calculating with it in \cite{jorsha}, so its
apparent simplicity is deceptive.  Second, although Sudarshan has
always emphasized that the most general time evolution of a density
operator can always be cast in linear form, the form shown here is
instead affine, and that is the source of some of its charming
features.  Notice that the linear term depends only on the time
evolution operator $U$ and not on any correlations between the
subsystem and the environment, while the contribution from the
correlations has been isolated in the affine term.  Third, this means
that in fact $M$ is not a function of only $\Tr_\mathcal{E} A$, but
$A$ in its entirety, as of course it must be; the idea here is not to
neglect the correlations but to put them in a more tractable form.
Finally, since this expression is available generally, it clearly
follows that the $M(t_i, t_{i-1})$, while always defined, will not in
general form a semigroup (as this would imply a master equation), and
forming a group is almost always out of the question, as that would
imply Hamiltonian evolution of the subsystem.

Casting $M(t_i,t_{i-1})$ in this form makes one point particularly
clear.  While it is typically assumed that the time evolution of the
preferred subsystem obeys a master equation, is Markovian, or is
otherwise special, all of those assumptions are for computational
simplicity only; none of them are needed to express decoherence of
histories.  Stamp forcefully expresses a concern, hinted at previously
in this paper, that the work which has been done up to this point on
decoherence has not considered sufficiently general situations to
convince the general population of physicists that truly generic
mechanisms for producing the classical world have in fact been
unearthed \cite{stamp}; and while I don't wish to endorse Stamp's
position fully, I do agree that the model calculations performed up to
this point don't quite convey the flavor of a universal phenomenon.
This leads me to the fundamental point of this paper: I suggest that a
study of decoherence functionals using the formalism of Jordan
\textit{et al.}\ would give a broader and more general picture of
decoherence as a phenomenon than the sorts of calculations largely
performed up to this point have done.  Jordan \textit{et al.}\
state that the aim of their paper is ``to simply describe the
Schr\"{o}dinger picture before making approximations to it'' (such as
the introduction of master equations), and I suggest that such a 
perspective would benefit the study of decoherence, which in many ways
is so much more mature than it was a decade ago.

This maturity shows itself particularly in the research program of
Zurek and coworkers, recently summarized in \cite{zurek}, in which it
is recognized that decoherence is only one of many ingredients that
characterize classicality.  The larger picture is that the classical world
is characterized by particular observables that carry several properties:
\begin{itemize}
\item The corresponding eigenstates are highly resistant to environmental dephasing,
      while their superpositions are not.
\item Thus they are selected by the details of dynamics, not our arbitrary choices;
      it is dynamics that tells us what we can potentially observe.
\item These observables imprint their values highly redundantly on the environment,
      so it is possible to deduce their values by sampling only a small fraction of
      the environment (knowing the position of an object after sampling only very few
      of the photons reflected from it is an obvious example).
\item Finally, histories composed of projections onto eigenstates of these observables
      exhibit decoherence.
\end{itemize}
Many of these issues have been studied in explicit models in a master
equation context, and I suggest that further study using this more
general open systems perspective would be of great value.

However, the results of Jordan \textit{et al.}\ require extension in
certain directions to be useful here.  The most obvious is that
Eqs. (\ref{linpart}) and (\ref{affpart}) require the dimension of the
environment's Hilbert space to be finite, as they involve the
introduction of a ``state of complete ignorance'' of the environment,
which is defined only in the finite-dimensional case.  While this is adequate
for spin bath calculations, an infinite-dimensional version is clearly required
for other applications.  Perhaps this is easily done by replacing the state of
complete ignorance with a thermal equilibrium state, but that remains to be 
demonstrated.

\section{Conclusions}

Establishing that quantum mechanics has the correct classical limit
has proved to be a surprisingly challenging problem; some time passed
before the full extent of the problem was even clear.  Since the
recognition that decoherence was an important contributor, however,
various research programs have made progress on this issue.  Here I
have summarized results due to Griffiths, Omn\`{e}s, and Gell-Mann and
Hartle, and I have argued that ``decoherence'' or ``consistency'' in
their sense is necessary for establishing one aspect of the classical
limit, namely the validity of classical statistics.  I have also
pointed out that decoherence is actually only one ingredient among
many for demonstrating this validity in certain regimes.  After
describing their formalism, I tried to bring it into contact with
approaches to decoherence that consider not histories but reduced
density matrices of selected subsystems, and I found that the
connection seemed to depend on making certain approximations in the
quantum theory of open systems, such as the existence of Markovian
master equations.  I also suggested that this might cast some doubts
on claims that truly universal mechanisms explaining the emergence of
classicality have been found.  Then I described a way to analyze
histories based on recent work by Jordan, Shaji, and Sudarshan that
seems to justify this connection while avoiding the usual
approximations, thus offering the possibility of demonstrating
decoherence in a much broader range of situations.  I suggest that
this work offers a new path forward for performing calculations that
can illuminate many different aspects of the classical limit,
including but certainly not limited to decoherence of histories, with
considerably greater generality.

\ack 
Clearly I am indebted first and foremost to E.\ C.\ George Sudarshan,
to whom this paper is dedicated.  Knowing my interest in foundational
issues in quantum theory, he introduced me to Gell-Mann and Hartle's
early articles on decoherence, and our analysis of their work and
others' has influenced my understanding of these issues long after I
moved on to other topics in my day job.  I'd also like to acknowledge
Thomas Jordan, with whom Sudarshan and I wrote \cite{us}; he has
written many clear and useful articles on foundational issues in
quantum theory, and he continues to do so to this day.  I also
appreciate very helpful conversations with Wojciech Zurek, who changed
my views on some of the issues described here and also pointed me to
some important current references.  Finally, I should also mention
that this work was supported by the U.\ S.\ Department of Energy under
contract DE-AC52-06NA25396.


\section*{References}

\begin{thebibliography}{99}
\bibitem{griffiths} Griffiths R B 1984 \textit{Journal of Statistical Physics} 
                    \textbf{36} 219
\nonum Griffiths R B 1986 \textit{Fundamental Questions in Quantum Mechanics} 
       ed L M Roth and A Inomata (New York: Gordon and Breach)
\nonum Griffiths R B 1986 \textit{New Techniques and Ideas in Quantum 
       Measurement Theory} ed D M Greenberger (New York: New York Academy of 
       Sciences)
\nonum Griffiths R B 1987 \textit{American Journal of Physics} \textbf{55} 11
\nonum Griffiths R B 1993 \textit{Physical Review Letters} \textbf{70} 2201
\nonum Griffiths R B 1996 \textit{Physical Review} A \textbf{54} 2759
\bibitem{omnes} Omn\`{e}s R 1988 \textit{Journal of Statistical Physics} 
                \textbf{53} 893, 933, 957
\nonum Omn\`{e}s R 1989 \textit{Journal of Statistical Physics} \textbf{57} 357
\nonum Omn\`{e}s R 1990 \textit{Annals of Physics} \textbf{201} 354
\nonum Omn\`{e}s R 1991 \textit{Journal of Statistical Physics} \textbf{62} 841
\nonum Omn\`{e}s R 1994 \textit{The Intepretation of Quantum Mechanics} 
       (Princeton: Princeton University Press)
\bibitem{GMH} Gell-Mann M and Hartle J B 1990 \textit{Complexity, Entropy, and 
              the Physics of Information} (SFI Studies in the Sciences of 
              Complexity vol 8) ed W Zurek (Redwood City: Addison Wesley)
\nonum Gell-Mann M and Hartle J B 1990 \textit{Proceedings of the 3rd 
       International Symposium on the Foundations of Quantum Mechanics in the 
       Light of New Technology} ed S Kobayashi, H Ezawa, M Murayama, and S 
       Nomura (Tokyo: Physical Society of Japan)
\nonum Gell-Mann M and Hartle J B 1990 \textit{Proceedings of the 25th 
       International Conference on High Energy Physics} ed K K Phua and Y 
       Yamaguchi (Singapore: World Scientific)
\nonum Gell-Mann M and Hartle J B 1993 \textit{Physical Review} D \textbf{47} 
       3345
\nonum Gell-Mann M and Hartle J B 1994 \textit{Proceedings of the NATO Workshop 
       on the Physical Origins of Time Asymmetry} ed J J Halliwell, J 
       Perez-Mercader, and W Zurek (Cambridge: Cambridge University Press)
\bibitem{wig} Wigner E P 1963 \textit{American Journal of Physics} 
              \textbf{31} 6
\bibitem{ABL} Aharanov Y, Bergmann P G, and Lebowitz J L 1964 \textit{Physical 
              Review} \textbf{134} B1410
\bibitem{revmodphys} Omn\`{e}s R 1992 \textit{Reviews of Modern Physics} 
                     \textbf{64} 339
\bibitem{dowkent} Dowker F and Kent A 1995 \textit{Physical Review Letters} 
                  \textbf{75} 3038
\nonum Dowker F and Kent A 1996 \textit{Journal of Statistical Physics} 
       \textbf{82} 1575
\bibitem{us} Chisolm E, Sudarshan E C G, and Jordan T F 1996 
             \textit{International Journal of Theoretical Physics} \textbf{35} 
             485
\bibitem{choice} Griffiths R B 1998 \textit{Physical Review} A \textbf{57} 1604
\bibitem{grifbook} Griffiths R B 2003 \textit{Consistent Quantum Theory} 
                   (Cambridge: Cambridge University Press)
\bibitem{bell} Bell J S 1990 \textit{Physics World} \textbf{3} 33
\bibitem{stamp} Stamp P C E 2006 \textit{Studies in History and Philosophy of 
                Modern Physics} \textbf{37} 467
\bibitem{feynvern} Feynman R P and Vernon F L 1963 \textit{Annals of Physics} 
                   \textbf{24} 118
\bibitem{caldleg} Caldeira A O and Leggett A J 1983 \textit{Annals of Physics} 
                  \textbf{149} 374
\nonum Caldeira A O and Leggett A J 1983 \textit{Physica} A \textbf{121} 587
\nonum Caldeira A O and Leggett A J 1985 \textit{Physical Review} A 
       \textbf{31} 1059 
\bibitem{prosta} Prokof'ev N V and Stamp P C E 2000 \textit{Reports on 
                 Progress in Physics} \textbf{63} 669
\bibitem{dowhal} Dowker H F and Halliwell J J 1992 \textit{Physical Review} D 
                 \textbf{46} 1580
\bibitem{zurek} Zurek W H 2003 \textit{Reviews of Modern Physics} \textbf{75} 
                715
\bibitem{joos} Joos E, Zeh H D, Kiefer C, Giulini D, Kupsch J, and Stamatescu 
               I-O 2003 \textit{Decoherence and the Appearance of a Classical 
               World in Quantum Theory} 2nd ed (Berlin: Springer-Verlag)
\bibitem{paz} Paz J P and Zurek W H 1993 \textit{Physical Review} D 
              \textbf{48} 2728
\bibitem{jorsha} Jordan T F, Shaji A, and Sudarshan E C G 2006 
                 \textit{Physical Review} A \textbf{73} 012106
\end{thebibliography}
\end{document}